\documentclass[epj]{svjour}
\usepackage{amsmath}
\usepackage{graphicx}
\usepackage{subfigure}
\usepackage{epsfig}
\usepackage{dcolumn}
\usepackage{bm}

\begin{document}

\title{Total entropy production fluctuation theorems in a nonequilibrium time-periodic steady state}

\author{Sourabh Lahiri and A. M. Jayannavar\thanks{\email{jayan@iopb.res.in}}}

\institute{Institute of Physics, Sachivalaya Marg, Bhubaneswar-751005, India}

\date{\today}

\newcommand{\nwc}{\newcommand}
\nwc{\beq}{\begin{equation}}
\nwc{\eeq}{\end{equation}}
\nwc{\bdm}{\begin{displaymath}}
\nwc{\edm}{\end{displaymath}}
\nwc{\bea}{\begin{eqnarray}}
\nwc{\eea}{\end{eqnarray}}
\nwc{\para}{\paragraph}
\nwc{\vs}{\vspace}
\nwc{\hs}{\hspace}
\nwc{\la}{\langle}
\nwc{\ra}{\rangle}
\nwc{\del}{\partial}
\nwc{\lw}{\linewidth}
\nwc{\nn}{\nonumber}
\nwc{\pd}[2]{\frac{\partial #1}{\partial #2}}

\abstract{
 We investigate the total entropy production of a Brownian particle in a driven bistable system. This system exhibits the phenomenon of stochastic resonance. We show that in the time-periodic steady state, the probability density function for the total entropy production satisfies Seifert's integral and detailed fluctuation theorems over finite time trajectories.
}

\titlerunning{Total entropy production fluctuation theorem}
\maketitle{}

\section{Introduction}

Recent advances in non-equilibrium statistical mechanics address, for the first time, questions relevant to modelling of nanoscale machinery or to thermodynamics of small systems \cite{bus05}. Of particular interest are the so-called \emph{fluctuation theorems}\cite{eva02,har07,rit03,rit06,kur07,eva93,eva94,gal95,kur98,leb99,cro99,cro00,jar97,zon02,zon04a,abh04,sei}. They reveal rigorous relations for properties of distribution functions of physical variables such as work, heat and entropy production for systems driven away from equilibrium, where Einstein's and Onsager's relations no longer hold. Some of these theorems have been verified experimentally \cite{cil98,wan02,lip02,dou05,tre04,dou06,zon04b,spe07,bli06,jou08,pet08,imp08}. 
One of the fundamental laws of physics, the second law of thermodynamics, states that the entropy of an isolated system always increases. The second law is statistical in nature. At small time scales or in small systems, one can observe excursions away from the typical behaviour. The entropy production in a given time, being a fluctuating quantity, can take negative values and yet the average entropy production over all times is always positive. Entropy or entropy production is generally considered as an ensemble property. 
However, Seifert has generalized the concept of entropy to a single stochastic trajectory \cite{sei,sei05,imp06}. The total entropy production along a single trajectory involves both particle entropy and entropy production in the environment. The total entropy production is shown to obey the integral fluctuation theorem (IFT) for any initial condition and drive, over any finite time interval. In stationary state, a stronger fluctuation theorem, namely the detailed fluctuation theorem (DFT) holds. Applications of entropy production fluctuation theorems in different physical contexts can be found in references \cite{gom06,tie06,sch07,xia08}.

In the present work, we probe numerically the afore mentioned entropy production fluctuation theorems (the IFT and the DFT) in the case of a Brownian particle placed in a double well potential and subjected to an external harmonic drive. In the absence of drive, the particle hops between the two wells with Kramer's escape rate $r_k=\tau_0^{-1}e^{-\Delta V/k_BT}$ \cite{han90} where $\tau_0$ is a characteristic time, $\Delta V$ is the energy barrier height between the two symmetric wells and $T$ is the temperature of the bath.
 The random hops of the Brownian particle between the two wells get synchronized with the external drive if $r_k$ matches twice the frequency of the external drive. This optimization condition can be achieved by tuning the noise intensity, and is called \emph{stochastic resonance} (SR) \cite{ben82,gam98,gam95}. Noise plays a constructive role in this case and SR finds applications in almost all areas of natural sciences.
 To characterize this resonance behaviour, different quantifiers have been introduced in the literature \cite{gam98,gam95,mah97a,mah97b,evs05,mah98,iwa01,dan05}. The work injected into the system (or the thermodynamic work done on the system) per cycle characterizes SR as a bona fide resonance \cite{gam95,iwa01,dan05}. Recently work and heat fluctuation theorems have been analyzed in a symmetric double well system exhibiting SR in presence of external subthreshold harmonic \cite{sai07,sah08} and biharmonic \cite{sin08} drives. Theoretical \cite{sai07,sah08} and experimental \cite{jou08,pet08} studies reveal the validity of the steady state fluctuation theorem (SSFT) for heat and work integrated over finite time intervals. In the following, we extend the study to fluctuation theorems for total entropy production and associated probability density functions.

\section{The Model}

The overdamped dynamics for the position ($x$) of the particle is given by a Langevin equation \cite{ris} in a dimensionless form, namely

\beq
\frac{dx}{dt×} = -\pd{U(x,t)}{x} + \xi(t),
\eeq

where $\xi(t)$ is the Gaussian white noise with $\la \xi(t)\ra=0$ and $\la \xi(t)\xi(t')\ra=2D\delta(t-t')$, where the noise strength $D=k_BT$, $k_B$ being the Boltzmann constant. The potential $U(x,t)$ can be split into two parts: a static potential $V(x)=-\frac{1}{2×}x^2+\frac{1}{4×}x^4$, and the potential due to external harmonic perturbation $V_1(x,t)=-xA\sin\omega t$. $A$ and $\omega$ are amplitude and frequency of the external drive, respectively. 
The static double well potential $V(x)$ has a barrier height $\Delta V=0.25$ between two symmetrically placed wells (or minima) located at $x_m=\pm 1$. We have restricted our analysis to subthreshold forcings, $A|x_m| < \Delta V$. The total potential $U(x,t)=V(x)+V_1(x,t)$. Using the method of stochastic energetics \cite{sek97} for a given particle trajectory $x(t)$ over a finite time duration $\tau$, the physical quantities such as injected work or thermodynamic work ($W$), change in internal energy ($\Delta U$) and heat ($Q$) dissipated to the bath can be calculated. They are given by

\begin{subequations}
\beq
W=\int_{t_0}^{t_0+\tau}\pd{U(x,t)}{t}dt, ~~~~~~~~~~~~~~~~~~~~~~
\eeq

\beq
\hspace{1cm}\Delta U = U(x(t_0+\tau),t_0+\tau)-U(x(t_0),t_0), \mbox{~~and}
\eeq

\beq
 Q=W-\Delta U. ~~~~~~~~~~~~~~~~~~~~~~~~~~~~~~  
\label{2c}
\eeq
\end{subequations}

Equation (\ref{2c}) is a statement of the first law of thermodynamics. The particle trajectory extends from initial time $t_0$ to final time $t_0+\tau$. $W$, $\Delta U$ and $Q$ are all stochastic quantities and we have evaluated them numerically by solving Langevin equation using Heun's method \cite{sai07,sah08,man00}.

A change in the medium entropy ($\Delta s_m$) over a time interval $\tau$ is given by

\beq
\Delta s_m = \frac{Q}{T}.
\label{3}
\eeq

The nonequilibrium entropy $S$ of the system is defined as 

\beq
S(\tau) = -\int dx ~P(x,t) \ln P(x,t) = \la s(\tau)\ra.
\label{4}
\eeq

Using (\ref{4}), Seifert defines a trajectory dependent entropy of the particle as \cite{sei,sei05}

\beq
s(t)=-\ln P(x,t),
\label{5}
\eeq

where $P(x,t)$ is obtained by solving the dynamical equation for probability density evaluated along the stochastic trajectory $x(t)$. The change in the system entropy for any trajectory of duration $\tau$ is given by

\beq
\Delta s = -\ln \left[ \frac{P_1(x(t_0+\tau),t_0+\tau)}{P_0(x(t_0),t_0)} \right],
\label{6}
\eeq

where $P_0(x(t_0),t_0)$ and $P_1( x(t_0+\tau),t_0+\tau)$  are the probability densities of the particle positions at initial time $t_0$ and final time $t_0+\tau$ respectively. Thus for a given trajectory $x(t)$, the change in entropy $\Delta s$ depends on the initial probability density and hence contains the information about the whole ensemble. The total entropy change over time duration $\tau$ is given by

\beq
\Delta s_{tot} = \Delta s_m + \Delta s.
\label{7}
\eeq

Using the above definition of total entropy production, Seifert has derived the IFT \cite{sei,sei05}, i.e.,

\beq
\la e^{-\Delta s_{tot}}\ra = 1,
\label{8}
\eeq

where angular brackets denote average over the statistical ensemble of realizations, i.e., over the ensemble of finite time trajectories.

This identity is very general and holds at any time and for arbitrary initial conditions. Equation (\ref{8}) along with Jensen's inequality implies $\la\Delta s_{tot}\ra\ge 0$ which is a refined interpretation of the second law. This still leaves open the possibility that there exists individual realizations for which this $\Delta s_{tot}$ is negative. In the presence of external periodic perturbations, the system relaxes to a time-periodic steady state. In this state, a stronger detailed fluctuation theorem holds \cite{cro99,sei,sei05}:

\beq
\frac{P(\Delta s_{tot})}{P(-\Delta s_{tot})} = e^{\Delta s_{tot}},
\label{9}
\eeq

where $\Delta s_{tot}$ is evaluated over time intervals $\tau=nT_1$, $T_1$ being the period of the external drive. $P(\Delta s_{tot})$ (or $P(-\Delta s_{tot})$) is the probability that the trajectory produces (or consumes) entropy with the magnitude $\Delta s_{tot}$.

To calculate the total entropy production, we evolve the Langevin system under the time-periodic force over many realizations of noise. Ignoring transients, we first find out probability density function  $P(x,t)$ in the time asymptotic regime. In this case, $P(x,t)$ is a periodic function in $t$ with the period equal to that of the external drive. Having evaluated the time-periodic probability density function, we again evolve the system trajectory. The heat dissipated is calculated over a period (or over a number of periods) using (\ref{2c}). 
Thereby we obtain the change in the medium entropy ($\Delta s_m=Q/T$).
Knowing the end-points of each trajectory, and the time-periodic $P(x,t)$, the change in system entropy $\Delta s$ is calculated (equation (\ref{6})). Thus we obtain for each trajectory the total entropy production ($\Delta s_{tot}=\Delta s_m+\Delta s$) over a single trajectory. To calculate the averages of the physical quantities or the probability distribution, $\Delta s_{tot}$ is obtained for more than $10^5$ realizations. In the following we present the results where all the physical parameters are taken in dimensionless form.

\section{Results and Discussions}

The work $W_p$  calculated over a period chosen at random varies from realization to realization and is a random quantity. So are $U_p$ and $Q_p$. However, all these quantities satisfy equation \ref{2c} for each period chosen at random. The averages of physical quantities ($\la\cdots\ra$) are calculated over $10^5$ realizations.

\begin{figure}
\vspace{0.5cm}
\centering
 \epsfig{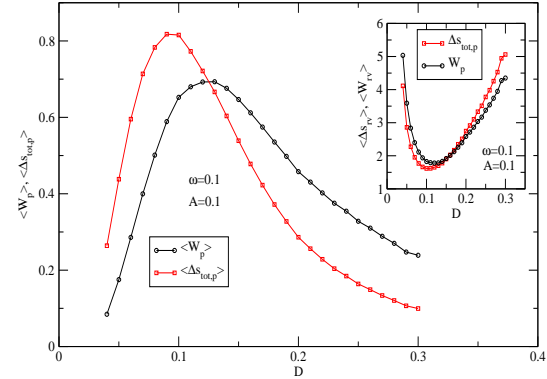}
\caption{Variation of $\la W_p\ra (=\la Q_p\ra)$ and $\la \Delta s_{tot,p}\ra$ with $D$, for $A=0.1, ~\omega=0.1$. The curve of $\la W_p\ra$ versus $D$ has been scaled by a factor of 8 for clarity. Inset shows the corresponding relative variances as a function of $D$.}
\label{f1}
\end{figure}

In figure \ref{f1}, we have plotted the average work done (or injected work) $\la W_p\ra$ over a single period of the external drive in  nonequilibrium time-periodic state, as a function of $D$ for $A=0.1$. The internal energy $U$ being a state variable, $\la U_p\ra$ is periodic in time and hence $\la\Delta U_p\ra=0$. From equation (\ref{2c}) we find that the average heat dissipated over a period $\la Q_p\ra$ equals the average work $\la W_p\ra$ done over a period.
In the same figure, average total entropy production over a single period, $\la\Delta s_{tot,p}\ra$, as a function of $D$ has also been plotted. 
Since entropy of system is a state variable, $\la\Delta s_{tot,p}\ra=\la\Delta s_{m,p}\ra=\la\frac{Q_p}{T}\ra$. The average work or heat exhibits a well-known SR peak (around $D=0.12$) consistent with the condition (at low frequency of drive) of matching between Kramer's rate and frequency of drive, which has been studied in earlier results \cite{iwa01,dan05}. However,peak in the $\la\Delta s_{tot,p}\ra$ is not at the same $D$ at which SR condition is satisfied. It is expected that at resonance, system will absorb maximum energy from the medium and being in a stationary state, will release this same energy back to the medium. 
The peak for $\la\Delta s_{tot,p}\ra$ not being at the same temperature as that for $\la Q_p\ra$ or $\la W_p\ra$ is understandable as $\la\Delta s_{tot,p}\ra=\la\Delta s_{m,p}\ra=\la Q_p/T\ra$, i.e., peak in $\la Q_p\ra$ versus $T$ will be shifted if we plot $\la Q_p/T\ra$ versus $T$. 
[Similar observations are noted in the nature of directed current in ratchet systems. In these periodic systems, unidirectional currents can be obtained in a nonequilibrium state in the absence of obvious bias. The average current exhibits a resonance peak as a function of temperature. Even though currents in these systems are generated at the expense of entropy, the peak in the total entropy production is not at the peak for the current.\cite{kri05}]
 In the inset of figure \ref{f1}, we have plotted the relative variance of work $\left(\la W_{rv}\ra \equiv \frac{\sqrt{\la W_{p}^2\ra-\la W_{p}\ra^2 }}{\la W_{p}\ra}\right)$ and that of total entropy $\left(\la\Delta s_{rv}\ra \equiv \frac{\sqrt{\la\Delta s_{tot,p}^2\ra-\la\Delta s_{tot,p}\ra^2 }}{\la\Delta s_{tot,p}\ra}\right)$. $\la W_{rv}\ra$ exhibits a minimum around SR condition.
However, $\la\Delta s_{rv}\ra$ shows a minimum around the same temperature at which $\la\Delta s_{tot,p}\ra$ shows a peak. Thus, unlike $\la W_{rv}\ra$ \cite{imp08,sai07,sah08}, $\la\Delta s_{rv}\ra$  cannot be used as a quantifier of SR. 
This is because the minimum in $\la\Delta s_{rv}\ra$ is correlated to the peak in $\la \Delta s_{tot}\ra$ as a function of $D$, which itself does not occur at the value of the $D$ at which resonance condition is satisfied, as discussed earlier.
It may be noted that relative variance of both work and total entropy production over single period are larger than 1, implying that these quantities are not self-averaging (i.e., fluctuation dominates the mean). However, when the observation time for the stochastic trajectory is increased to a large number ($n$) of periods, the relative variance, which scales as $n^{-1/2}$, becomes a self-averaging quantity, i.e., mean is larger than the dispersion \cite{sah08}.

\vspace{0.2cm}

\begin{figure}
\vspace{0.5cm}
\centering
\epsfig{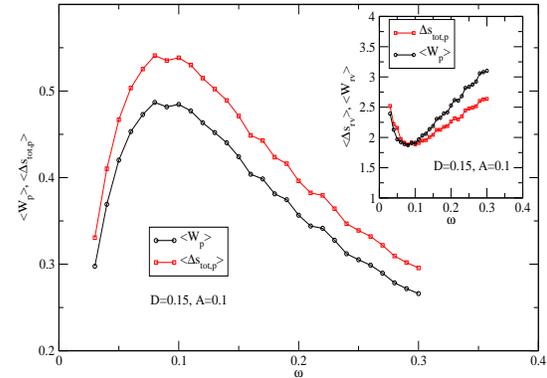}
\caption{Variation of $\la W_p\ra=\la Q_p\ra$ and $\la \Delta s_{tot,p}\ra$ with $\omega$, for $A=0.1, ~D=0.15$. The curve of $\la W_p\ra$ versus $\omega$ has been scaled by a factor of 6 for easy comparison. Inset shows the corresponding relative variances as a function of $\omega$.}
\label{f2}
\end{figure}

In figure \ref{f2}, we have plotted $\la W_p\ra$ and $\la\Delta s_{tot,p}\ra$ as a function of $\omega$. The injected work $\la W_p\ra$, exhibits a peak as a function of $\omega$, thus characterizing SR as a bona fide resonance \cite{gam95,iwa01,dan05}. It may be noted that the peak position for $\la \Delta s_{tot,p}\ra$, in this case, is at the same value as that for $\la W_p\ra$ or $\la Q_p\ra$, as expected. The inset shows the relative variance of $\Delta s_{tot}$ versus frequency of external drive $\omega$ which in turn shows a minimum at the resonance condition.

\begin{figure}
\centering
 \begin{tabular}{c}
\epsfig{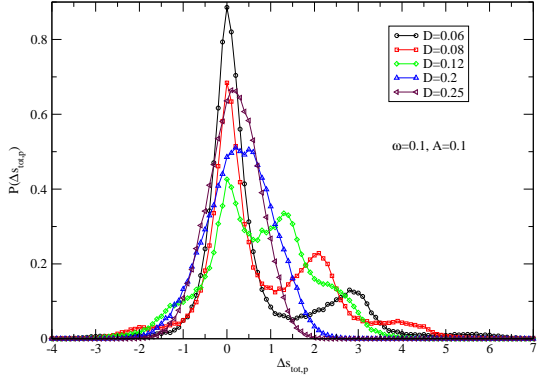} 
\end{tabular}
\caption{Plots of probability distribution functions of $\Delta s_{tot,p}$ for different values of noise strength $D$. The fixed parameters are: $A=0.1, ~\omega=0.1$.}
\label{f3}
\end{figure}

In figure \ref{f3}, we have plotted the probability distribution $P(\Delta s_{tot,p})$ versus $\Delta s_{tot,p}$, for different values of noise strength spanning a region of temperatures around that of SR ($D=0.12$). For low temperature side, $D=0.06$, $P(\Delta s_{tot,p})$ exhibits a double peak structure. The peak around zero can be attributed to the intrawell motion. The small peak at higher positive values of $\Delta s_{tot,p}$ is caused by the occasional interwell transition which entails larger heat dissipation in the medium and contributes to the total entropy production via entropy produced in the bath, $\la\Delta s_{m,p}\ra = \la Q_p\ra/T$. 
At very low temperature, $D=0.02$, the interwell motion is subdominant (particle exhibits small oscillations about the minimum). $P(\Delta s_{tot,p})$ exhibits a single peak around $\la\Delta s_{tot,p}\ra$ and the distribution is closer to Gaussian which is not shown in the graphs. As temperature is increased, due to the enhancement of interwell motion, peak at the right increases. These multipeaked distributions are asymmetric. The distributions extend to the negative side. Finite values of distributions in the negative side is necessary to satisfy fluctuation theorems. 
The contribution to the negative side comes from the trajectories which lead to transient violations of the second law. For higher values of temperature, $D=0.25$ (and beyond), the peak structures merge and $P(\Delta s_{tot,p})$ becomes closer to a Gaussian distribution. Similar observations have been made for distributions of work and heat in earlier literature \cite{sai07,sah08}. The observed values of $\la e^{-\Delta s_{tot,p}}\ra$, from our simulations, are equal to 1.045, 1.017, 0.980, 1.024 and 1.032, for values of temperatures $D=0.06, 0.08, 0.12, 0.2$ and 0.25, respectively. All the values for $\la e^{-\Delta s_{tot,p}}\ra$ are close to unity within our numerical accuracy, which is clearly consistent with IFT (equation (\ref{8})).

\begin{figure}
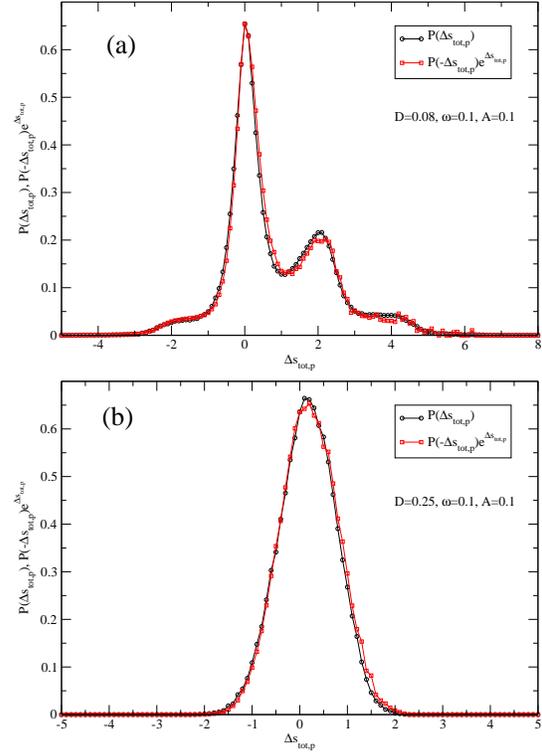

\centering
 \begin{tabular}{c}
\epsfig{file=fig4a.eps,width=0.8\lw,clip=}\\
\epsfig{file=fig4b.eps,width=0.8\lw,clip=} 
\end{tabular}
\caption{(a) Both $P(\Delta s_{tot,p})$ and $P(-\Delta s_{tot,p})e^{\Delta s_{tot,p}}$ have been plotted on the same graph for $D=0.08, ~\omega=0.1$ and $A=0.1$. These curves match to a good accuracy, thereby providing a cross-verification for the validity of DFT. (b) Similar plots for $D=0.25$. Other parameters are the same as in (a).}
\label{f4}
\end{figure}

We have plotted  $P(\Delta s_{tot,p})$ and  $P(-\Delta s_{tot,p})e^{\Delta s_{tot,p}}$ on the same graph for two values of $D$ ($D=0.08$ and 0.25) in figures \ref{f4}(a) and (b) respectively, which abides by equation (\ref{9}), namely the DFT. We would like to mention that the IFT and DFT are exact theorems for a driven Langevin system. Our results corresponding to figures \ref{f4}(a) and (b) act as a check on the quality of our simulation.

\begin{figure}
\vspace{0.3cm}
\centering
 \epsfig{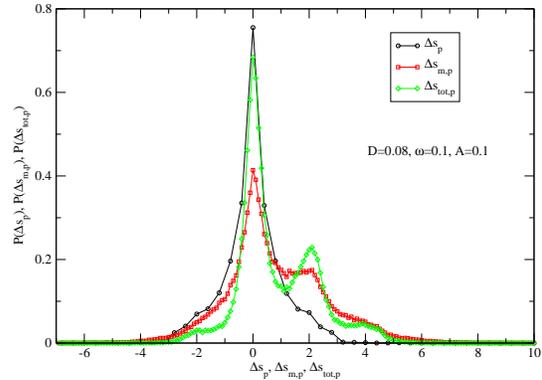}
\caption{Plots showing distribution functions of $\Delta s_{tot,p}$ (green), $\Delta s_{m,p}$ (red) and of $\Delta s_p$ (black), for $D=0.08$, $\omega=0.1$ and $A=0.1$.}
\label{f5}
\end{figure}

In figure \ref{f5}, we have plotted probability distributions of changes in total entropy $\Delta s_{tot,p}$, medium entropy $\Delta s_{m,p}$ and system entropy  $\Delta s_p$ over a single period for the parameter values $D=0.08, ~\omega=0.1$ and $A=0.1$. 
System entropy $s_p(t)$ is a state function and its average value is a periodic function of time in the asymptotic regime. Thus average change in the system entropy over a period is zero. 
Moreover, $P(\Delta s_p)$ is a symmetric function of $\Delta s_p$. The medium entropy is related to the heat dissipated along the trajectory ($\Delta s_{m,p}=Q_p/T$). The nature of $P(\Delta s_{m,p})$ is identical to that of heat distribution \cite{sah08}. 
All these probabilities exhibit finite contribution to the negative side. 

\begin{figure}
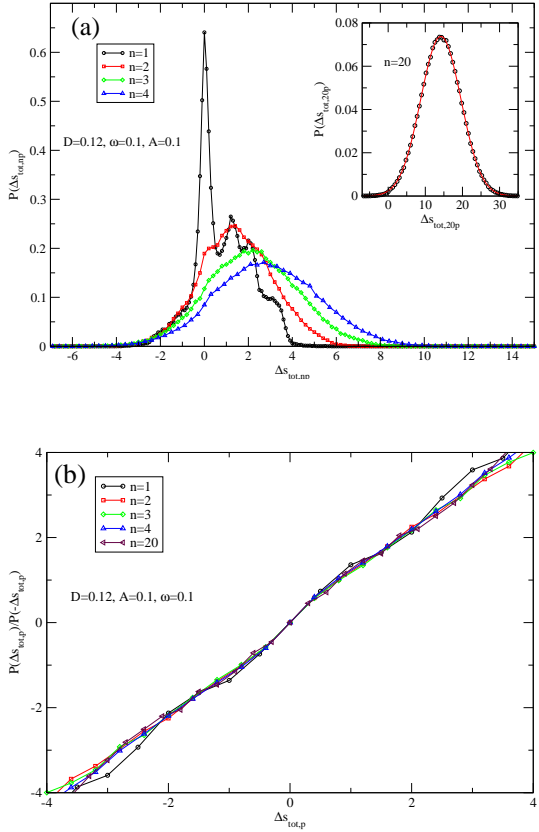

\vspace{0.3cm}
\centering
 \begin{tabular}{c}
\epsfig{file=fig6a.eps,width=0.8\lw,clip=}\\
 \\ \\
\epsfig{file=fig6b.eps,width=0.8\lw}
\end{tabular}
\caption{(a) Distributions of total entropy for different numbers of periods ($n=1, 2, 3$ and 4). The inset shows data points for $n=20$ and the corresponding Gaussian fit with $\sigma^2=28.61$ and $\la \Delta s_{tot,20p}\ra=14.18$. Parameter values are: $A=0.1, ~\omega=0.1$ and $D=0.12$. (b) Corresponding plots of symmetry functions of total entropy as a function of total entropy.}
\label{f6}
\end{figure}

As the observation time of the trajectory increases, there will be decrease in the number of trajectories for which $\Delta s_{tot}<0$. This is expected as we go to macroscopic scale in time. 
To this end we have plotted in figure \ref{f6}(a) the $P(\Delta s_{tot,np})$ obtained over different numbers (n) of cycles (or for observation times $\tau=nT_1$, where $T_1$ is the period of external drive). 
For a fixed value of the parameters, $D=0.12, A=0.1$ and $\omega=0.1$, and over single cycle, $P(\Delta s_{tot,p})$ exhibits multi-peaked structure which slowly disappears as we increase the period of observation. For larger periods, $P(\Delta s_{tot,np})$ tends closer to being a Gaussian distribution with a non-zero positive mean $\la\Delta s_{tot}\ra$. We also notice that as the number of periods increases, weight of the probability distributions to the negative side decreases.
In the inset of figure \ref{f6}(a), we have plotted probability density of $\Delta s_{tot}$ taken over 20 periods. The Gaussian fit is shown. The calculated values of variance, $\sigma^2=28.61$, and of the mean, $\la\Delta s_{tot}\ra=14.18$, closely satisfy the condition $\sigma^2=2\la\Delta s_{tot}\ra$, thereby abiding by the fluctuation-dissipation relation (see equation (18) of \cite{rit03}). If the distribution is a Gaussian and it satisfies the DFT, then the fluctuation-dissipation theorem $\sigma^2=2\la\Delta s_{tot}\ra$ must be satisfied \cite{rit03,cro99,jay08}.
The presence of long-time tails at large values of $\Delta s_{tot,np}$ are not ruled out (non-Gaussian nature of distribution). However, numerically it is difficult to detect them.

\begin{figure}
\vspace{0.5cm}
\centering
\begin{tabular}{c}
 \epsfig{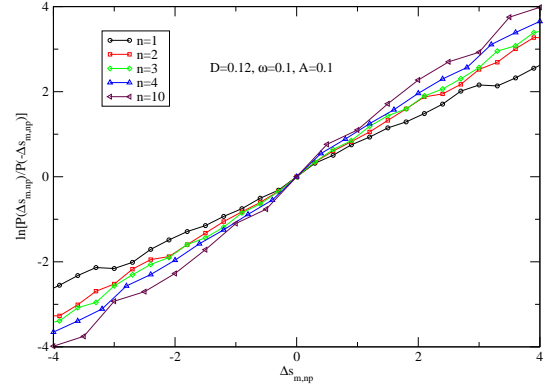}
\end{tabular}
\caption{The figure gives the plots of symmetry functions of medium entropy as a function of the medium entropy, for different numbers of periods. Parameter values are: $A=0.1, ~\omega=0.1$ and $D=0.12$.}
\label{f7}
\end{figure}

In figure \ref{f6}(b), we have plotted the symmetry functions $\left(\ln\left[ \frac{P(\Delta s_{tot,np})}{P(-\Delta s_{tot,np})}\right]\right)$ versus $\Delta s_{tot,np}$ for different periods. 
Irrespective of the number of periods, we find that slopes of all the curves are equal to 1, which is consistent with DFT. The validity of DFT implies IFT, but not vice versa. The medium entropy is extensive in time while the system entropy is not. Only over larger number of periods, the contribution to $\Delta s_{tot,np}$ from $\Delta s_{np}$ becomes very small as compared to $\Delta s_{m,np}$. This means that only over large time periods, $\Delta s_{m,np}$ obeys a DFT relation or steady state fluctuation theorem as noted in the earlier literature \cite{pet08,sah08} (see figure \ref{f7}). 
In this figure we have plotted symmetry functions for the medium entropy for different numbers of periods ($n$). As we increase $n$, the slope increases towards 1 and hence satisfies the DFT for large $n$. The value of $n$ over which $\Delta s_{m,np}$ follows DFT depends sensitively on the physical parameters, unlike the DFT for $\Delta s_{tot,np}$.

SR effect is detectable even in the presence of a supra-threshold signal \cite{dan05,apo97}, i.e., when $A|x_m|>\Delta V$. Our studies in this regime for the total entropy production reveal that the distributions are broad and asymmetric. Moreover the multipeak structures as observed in the subthreshold regime are absent.

\section{Conclusions}

In conclusion, we have studied the entropy production of a Brownian particle in a driven double well system which exhibits stochastic resonance. Average total entropy production per cycle shows a peak as a function of noise strength. However, it cannot be directly correlated to stochastic resonance condition.  Moreover, as the period of observation increases, contribution of negative total entropy producing trajectories decreases. In this nonlinear system, we have also analyzed these theorems in the transient regime. In this case, we obtain a rich structure for the probability distribution of trajectory dependent total entropy production. These results along with analytically solvable models will be published elsewhere \cite{lah08}.


\section{Acknowledgment}

One of us (AMJ) thanks DST, India for financial support.


\begin{thebibliography}{100}

\bibitem{bus05} C.~Bustamante,~J.~Liphardt  and  F.~Ritort,  Physics Today  {\bf 58}, 45 (2005).

\bibitem{eva02}  D.~J.~Evans  and  D.~J.~Searls,   Adv.Phys.  {\bf 51}, 1529 (2002).

\bibitem{har07} R.~~J.~~Harris and G.~~M.~~Sch\"{u}tz, J. Stat. Mech., p07020
(2007).

\bibitem{rit03} F. Ritort, Sem. Poincare {\bf 2} (2003) 63.

\bibitem{rit06} F. Ritort, J. Phys. Condens. Matter {\bf 18}, R531 (2006). 

\bibitem{kur07} J. Kurchan, J. Stat. Mech, p07005 (2007).
 
\bibitem{eva93} D. J. Evans, E. G. D. Cohen and G. P. Morris, Phys. Rev. Lett. {\bf 71}, 2401 (1993); {\bf 71}, 3616 (1993) [errata].

\bibitem{eva94} D. J. Evans and D. J. Searls, Phys. Rev. E {\bf 50}, 1645 (1994).

\bibitem{gal95} G. Galvotti and E. G. D. Cohen, Phys. Rev. Lett. {\bf 74}, 2694 (1995); J. Stat. Phys. {\bf 80}, 31 (1995).

\bibitem{kur98} J. Kurchan, J. Phys. A: Math. Gen. {\bf 31}, 3719 (1998).

\bibitem{leb99} J. L. Lebowitz and H. Spohn, J. Stat. Phys. {\bf 95}, 333 (1999).

\bibitem{cro99} G. E. Crooks, Phys. Rev. E {\bf 60}, 2721 (1999).

\bibitem{cro00} G. E. Crooks, Phys. Rev. E {\bf 61}, 2361 (2000).


\bibitem{jar97} C. Jarzynski, Phys. Rev. Lett. {\bf 78}(1997) 2690; Phys. Rev.~ E  {\bf 56} (1997) 5018.

\bibitem{zon02} R.~van~Zon  and E.~G.~D.~Cohen,  Phys.Rev. E  {\bf 67}  (2002) 046102

\bibitem{zon04a} R.~van~Zon  and E.~G.~D.~Cohen,  Phys.Rev. E  {\bf 69} (2004) 056121.

\bibitem{abh04} O.~Narayan and A.~Dhar, J. Phys.A:Math Gen {\bf 37}, 63 (2004).

\bibitem{sei} U. Seifert, Eur. Phys. J. B. {\bf 64}, 423 (2008).

\bibitem{cil98} S. Ciliberto and C. Laruche, J. Phys. IV France {\bf 8}, 215 (1998). 

\bibitem{wan02} G. M. Wang, E. M. Sevick, E. Mittag, D. J. Searls and D. J. Evans, Phys. Rev. Lett. {\bf 89}, 050601 (2002).

\bibitem{lip02} J. Liphardt, S. Dumont, S. B. Smith, I. Tinoco Jr., and C. Bustamante, Science {\bf 296}, 1832 (2002).

\bibitem{dou05} A. Petrosyan, F. Douarche, I. Rabbiosi and S. Ciliberto, Europhys. Lett. {\bf 70}, 593 (2005).

\bibitem{tre04} E. H. Trepagnier, C. Jarzynski, F. Ritort, G. E. Crooks, C. J. Bustamante and J. Liphardt, Proc. Natl. Acad. Sci. {\bf 101}, 15038 (2004).

\bibitem{dou06} F. Douarche, S. Joubaud, N. B. Garnier, A. Petrosyan and S. Ciliberto, Phys. Rev. Lett. {\bf 97}, 140603 (2006).

\bibitem{zon04b} R. von Zon, S. Ciliberto and E. G. D. Cohen, Phys. Rev. Lett. {\bf 92}, 130601 (2004).


\bibitem{spe07} T. Speck, V. Blickle, C. Bechinger and U. Seifert, Europhys. Lett. {\bf 79}, 30002 (2007).

\bibitem{bli06} V. Blickle, T. Speck, L. Helden, U. Seifert and C. Bechinger, Phys. Rev. Lett. {\bf 96}, 070603 (2006).

\bibitem{jou08} S. Joubaud, N. B. Garnier and S. Ciliberto, Europhys. Lett. {\bf 82}, 30007 (2008).

\bibitem{pet08} A. Petrosyan, P. Jop and S. Ciliberto, Europhys. Lett.{\bf 81} 50005 (2008).


\bibitem{imp08} A. Imparato, P. Jop, A. Petrosyan and S. Ciliberto, J. Stat. Mech. P10017 (2008).


\bibitem{sei05} U. Seifert, Phys. Rev. Lett. {\bf 95}, 040602 (2005).


\bibitem{imp06} A. Imperato, L. Peliti, Phys. Rev. E {\bf 74}, 026106 (2006).

\bibitem{gom06} A. Gomez-Marin and I. Pagonabarraga, Phys. Rev. E {\bf 74}, 061113 (2006).

\bibitem{tie06} C. Tietz, S. Schuler, T. Speck, U. Seifert and J. Wrachtrup, Phys. Rev. Lett. {\bf 97}, 050602 (2006).

\bibitem{sch07} T. Schmiedl, T. Speck and U. Seifert, J. Stat. Phys. {\bf 128}, 77 (2007).

\bibitem{xia08} T. Xiao, Z. Hou and H. Xin, J. Chem. Phys. {\bf 129}, 114506 (2008).


\bibitem{han90} P. Hanggi, P. Talkner, M. Borkovec, Rev. Mod. Phys. {\bf 62}, 251 (1990).

\bibitem{ben82} R. Benzi, G. Parisi, A. Sutera and A. Vulpiani, Tellus {\bf 34}, 10 (1982).

\bibitem{gam98} L. Gammaitoni, P. Hanggi, P. Jung and F. Marchesoni, Rev. Mod. Phys. {\bf 70}, 223 (1998).


\bibitem{gam95} L. Gammaitoni, F. Marchesoni and S. Santucci, Phys. Rev. Lett. {\bf 74}, 1052 (1995).


\bibitem{mah97a} M. C. Mahato and A. M. Jayannavar, Phys. Rev. E {\bf 55}, 6266 (1997).

\bibitem{mah97b} M. C. Mahato and A. M. Jayannavar, Mod. Phys. Lett. B {\bf 11}, 815 (1997).

\bibitem{evs05} M. Evstigneev, P. Riemann and C. Bechinger, J. Phys. C {\bf 17}, S3795 (2005).

\bibitem{mah98} M. C. Mahato and A. M. Jayannavar, Physica A {\bf 248}, 138 (1998).

\bibitem{iwa01} T. Iwai, Physica A {\bf 300}, 350 (2001).

\bibitem{dan05} D. Dan and A. M. Jayannavar, Physica A {\bf 345}, 404 (2005).

\bibitem{sai07} Shantu Saikia, Ratnadeep Roy and A.M. Jayannavar, Phys. Lett. A {\bf 369}, 367 (2007).

\bibitem{sah08} Mamata Sahoo, Shantu Saikia, Mangal C. Mahato, A.M. Jayannavar, Physica A {\bf 387}, 6284 (2008).


\bibitem{sin08} Navinder Singh, Sourabh Lahiri and A. M. Jayannavar, cond-mat/0806.4567.

\bibitem{ris} H. Risken, {\em The Fokker-Planck Equation: Methods of Solution and Applications} (Springer-Verlag Berlin, 1989).

\bibitem{sek97} K.~~Sekimoto,~~J.~~Phys.~~Soc.~~Jpn. {\bf 66} (1997)6335.

\bibitem{man00} R. Mannela, in: J.A. Freund and T. Poschel (Eds), {\em Stochastic Process in Physics, Chemistry and Biology, Lecture Notes in Physics}, vol. 557 Springer-Verlag, Berlin (2000) p353.

\bibitem{kri05} Raishma Krishnan and A. M. Jayannavar, Physica A {\bf 345}, 61 (2005).

\bibitem{jay08} A. M. Jayannavar and Mamata Sahoo, Phys. Rev. E {\bf 77}, 022105 (2008).

\bibitem{lah08} Arnab Saha, Sourabh Lahiri and A. M. Jayannavar, manuscript under preparation.


\bibitem{apo97} F. Apostolico, L. Gammaitoni, F. Marchesoni and S. Santucci, Phys. Rev. E {\bf 55}, 36 (1997).

\end{thebibliography}
\end{document}